\pdfoutput=1
\documentclass[%
 reprint,
 amsmath,amssymb,
 aps,
]{revtex4-2}

\usepackage{graphicx}
\usepackage{dcolumn}
\usepackage{bm}

\usepackage[utf8]{inputenc}
\usepackage[T1]{fontenc}
\usepackage{mathptmx}
\usepackage[T1]{fontenc} 
\usepackage{amsmath,color}
\usepackage{amssymb}
\usepackage{amsfonts}
\usepackage{amsthm}
\usepackage{mathtools}
\usepackage[title]{appendix}
\usepackage{hyperref}
\usepackage{braket}
\usepackage{xcolor}
\usepackage{etoolbox}
\usepackage{soul}
\newcommand*{\citen}[1]{%
  \begingroup
    \romannumeral-`\x 
    \setcitestyle{numbers}%
    \cite{#1}%
  \endgroup
}

\usepackage{algorithm}
\usepackage{algpseudocode}
\usepackage{dsfont}

\usepackage{bbold}

\begin{document}
\preprint{AIP/123-QED}

\title{Rectification of vibrational energy transfer in driven chiral molecules  }
\author{Jichen Feng}
\email{jcfeng@sas.upenn.edu}
\affiliation{ 
Department of Chemistry, University of Pennsylvania, Philadelphia, Pennsylvania 19104, USA}
\affiliation{Department of Chemistry, Princeton University, Princeton, New Jersey 
 08544, USA}
\author{Ethan Abraham}
\email{ethana@mit.edu}
\affiliation{ 
Department of Chemistry, Massachusetts Institute of Technology, Cambridge, Massachusetts 02139, USA}
\author{Joseph E. Subotnik}
\email{subotnik@princeton.edu}
\affiliation{\mbox{
Department of Chemistry, University of Pennsylvania, Philadelphia, Pennsylvania 19104, USA}}
\affiliation{Department of Chemistry, Princeton University, Princeton, New Jersey 08544, USA}
\email{subotnik@princeton.edu}
\author{Abraham Nitzan}
\email{anitzan@sas.upenn.edu}
\affiliation{ 
Department of Chemistry, University of Pennsylvania, Philadelphia, Pennsylvania 19104, USA}
\affiliation{ 
School of Chemistry, Tel Aviv University, Tel Aviv 69978, Israel
}

\date{\today}

\begin{abstract}
We show that the combination of molecular chirality and phase-controlled driving can lead to rectification of vibrational energy transfer. We demonstrate this effect using classical models of (1) a single helical chain and (2) a more realistic model of polyethylene double helix. We examine the effect of the driving frequency, polarization, and temperature on this phenomenon. Notably, we find that the direction and magnitude of the observed directionality preference depend on the driving frequency and phase, and that the effect persists at room temperature. 
\end{abstract}

\maketitle

\section{Introduction}
The correlation between linear and angular momenta of charge or energy carriers in chiral media, often referred to as linear-angular momentum locking \cite{yan2023locking,yan2023locking2,yan2024lockingreview}, has been recognized as an important factor governing electronic and spin transport properties. More recently, attention has also shifted to such correlations for nuclear vibrations with angular momentum, often referred to as chiral phonons, which have been observed and analyzed in a variety of systems \cite{zhu2018phononexp,ueda2023np,wang2024summary,ishito2023phononexp,zhang2015theory,zhang2019theory,CISS_interface,science2023magnetic,gao2024exp,juraschek2022theory,science2023magnetic,zhu2018phononexp,gao2024exp,ueda2023np}. While most studies have focused on such motions in crystalline systems, our earlier studies have revealed that angular–linear momentum locking can arise not only in crystalline solids but also in isolated chiral molecules \cite{ethan2023quantifying,jichen2024quantifying}. Interest in chiral phonons has been further fueled by proposals implicating them mechanistically in the chiral-induced spin selectivity (CISS) effect \cite{Spin-Seedback,CISS_interface,Fransson2022temperature,Fransson2020Prb,bian2025phase}. It has also been suggested \cite{chen2022diodetheory} that this correlation between linear and angular momentum can lead to preferred directionality of energy transfer in such systems but, since phonons with positive and negative angular (and corresponding linear) momenta appear symmetrically a crystal spectral distribution, the actual realization of this phenomenon is not assured.

In this paper we describe analytical and numerical studies of directional energy transport in chiral molecules. Our primary finding is that the combination of molecular chirality and phase-controlled driving can lead to highly directionally asymmetric energy flow without any asymmetry in the system-bath couplings.

The relationship between heat transport and chirality has previously been analyzed from several perspectives; for example, theoretical analyses based on the Boltzmann transport equation suggest that thermal gradients in chiral solids can serve as a source of nuclear angular momentum whose directionality depends on the handedness of the chiral solid  \cite{therm_angular,off-diagonal}.  Our recent work has predicted a similar effect, showing that heat transport across chiral molecular chains induces nuclear angular momentum, and that the effect exhibits robust properties, persisting in the steady state in both harmonic and anharmonic model systems \cite{Feng2025Nuclear}. Directional energy transport under mechanical or optical driving is another intriguing manifestation of dynamical chirality that provides a novel route towards directional control of energy flow.  We exhibit this effect in two model systems described in previous work \cite{Feng2025Nuclear,ethan2023chain,ethan2023quantifying,chen2022diodetheory} (which the single chain model is enhanced with including next-nearest-neighbour and next-next-nearest-neighbour coupling), and we examine its dependence on various parameters. Notably, we find that the effect persists at room-temperature.

\section{Molecular Models} 
The models chosen for this study are (1) a single (generally anharmonic) helical chain \cite{Feng2025Nuclear,chen2022diodetheory} that we study also in the harmonic approximation, and (2) an anharmonic polyethylene double helix \cite{ethan2023chain,ethan2023quantifying,Feng2025Nuclear}. 

The first model (henceforth referred to as Model 1) is a chain of atomic sites defined by the Hamiltonian \cite{PhysRevE.56.877,Feng2025Nuclear}
    \begin{equation}\label{harmonic}
        \begin{aligned}
            H_1&=\sum^N_{j=1}\frac{\Vec{p}_j^2}{2m}+V_1\left(\{\Vec{x_j}\} \right)+V_2\left(\{\Vec{x_j}\} \right)+V_3\left(\{\Vec{x_j}\} \right)
        \end{aligned}
    \end{equation}
    \begin{equation}
        V_{\nu}\left(\{\Vec{x}_j \}\right)=\sum^{N-\nu}_{j=1}\frac{1}{2}K\left[|(\Vec{x}_j+\Vec{r}_j)-(\Vec{x}_{j+\nu}+\Vec{r}_{j+\nu})| -a_\nu\right]^2
    \end{equation}
where $N$ is the number of atoms in the chain, $m$ is the atomic mass (taken to be equal across all atoms), $\Vec{r_j}$ are the equilibrium positions of the atoms and $\vec{x_j}$ are the displacements of the atoms (note that this Hamiltonian, as well as the Hamiltonian of Model 2, Eq.~13, are defined for the bare system without any external forces and without coupling to any dissipative bath). The model defines a single-chain helix obtained as follows: Taking the $x$-axis as the direction of a given bond, the next bond is generated by rotating the given bond around the  $y$-axis by $60^\circ$ followed by rotation around the $z$-axis by $120^\circ$. The chirality is enforced by the structural dependent equilibrium bond lengths of nearest-neighbor $a_1$, next nearest-neighbors $a_2$ and $a_3$. The position of each atom can be expressed as
\begin{equation}
    \Vec{r}_j=\left(r_\perp\cos(2j\pi/3), r_\perp\sin(2j\pi/3),jr_\parallel\right)
\end{equation}
where
\begin{equation}
    r_\perp=\sqrt{3}a_1/6
\end{equation}
\begin{equation}
    r_\parallel=\sqrt{3}a_1/2.
\end{equation}
Then we have
\begin{equation}
    a_2=\sqrt{13}a_1/2
\end{equation}
\begin{equation}
    a_3=3\sqrt{3}a_1/2.
\end{equation}
Making the harmonic approximation \cite{Feng2025Nuclear} leads to
    \begin{equation}
        V_{\nu_h}\left(\{\Vec{x}_j \}\right)=\sum^{N-\nu}_{j=1}\frac{1}{2}(\Vec{x}_j-\Vec{x}_{j+\nu})^T\textbf{K}_{j,j+\nu}(\Vec{x}_j-\Vec{x}_{j+\nu})
    \end{equation} where,
    \begin{equation}
        \textbf{K}_{j,j+1}=\mathbf{R}_{z}^{-1}(2\pi j/3)\mathbf{R}_{y}^{-1}(\pi/3)\textbf{K}_{0}\mathbf{R}_{y}(\pi/3)\mathbf{R}_{z}(2\pi j/3),
    \end{equation}
 
    \begin{widetext}
\begin{equation}
    \begin{aligned}
        \textbf{K}_{j,j+2}=&\mathbf{R}_{z}^{-1}(2\pi j/3+\pi/3)\mathbf{R}_{y}^{-1}[\arctan(2\sqrt{3})]\textbf{K}_{0}
        \mathbf{R}_{y}[\arctan(2\sqrt{3})]\mathbf{R}_{z}(2\pi j/3+\pi/3),
    \end{aligned}
    \end{equation}
   \end{widetext} 
\begin{equation}
        \textbf{K}_{j,j+3}=\mathbf{R}_{y}^{-1}(\pi/2)\textbf{K}_{0}\mathbf{R}_{y}(\pi/2),
    \end{equation}
    with
    \begin{equation}\label{equation:K0}
         \textbf{K}_{0}=\begin{pmatrix}
            K & 0 & 0 \\
            0 & 0 & 0 \\
            0 & 0 & 0 \\
        \end{pmatrix}.
    \end{equation} 
Here,  $\mathbf{R}_{y}$, $\mathbf{R}_{z}$ denote rotations about the $y$ and $z$-axes, respectively. Note that the harmonic model used in Ref.~\citen{Feng2025Nuclear} is a simplified version of this Hamiltonian that includes only the nearest-neighbor interactions (Eq. 8, $\nu=1$). In this harmonic approximation the energy transport dynamics can be evaluated analytically.

\begin{figure}
\includegraphics[width=240 pt]{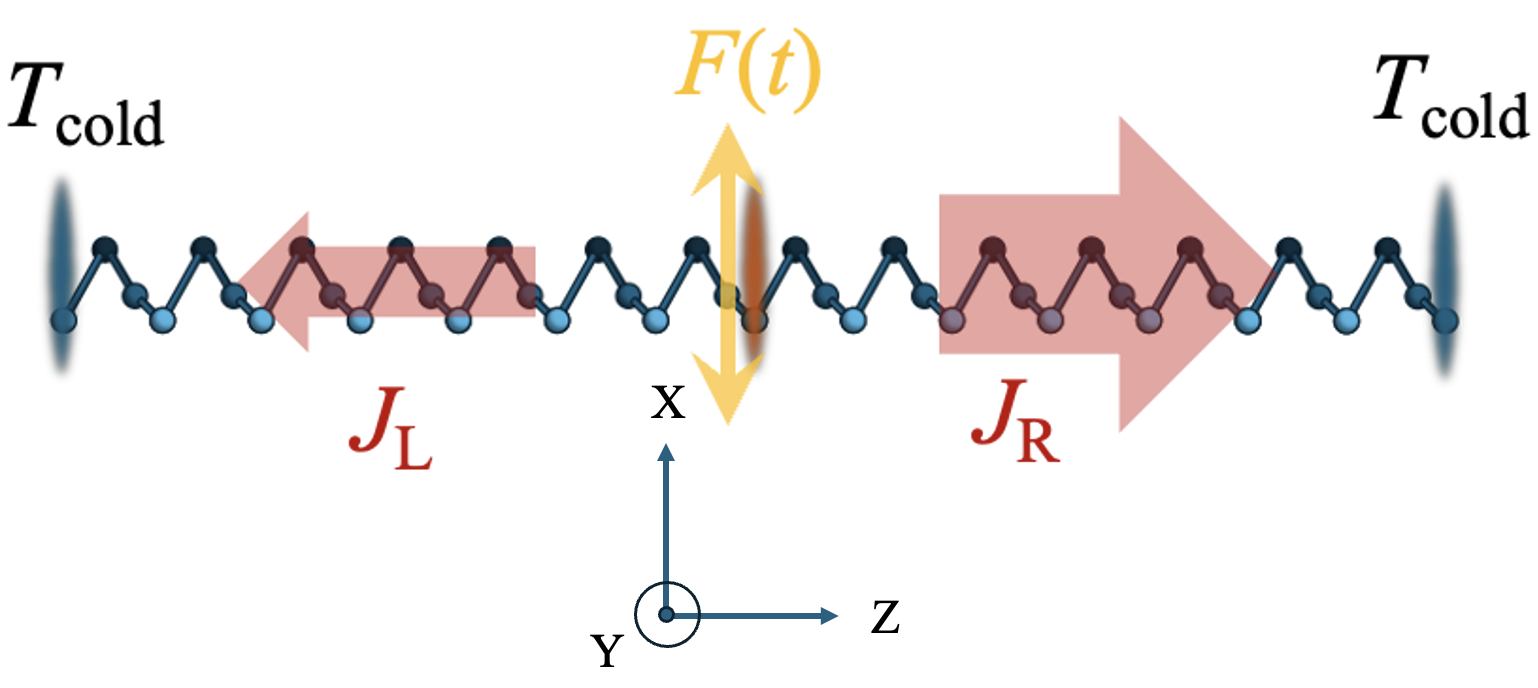}
\label{fig:1} 
\caption{Schematic of the construction used for this study: A chiral molecular chain is suspended in the $z$-direction by keeping the terminal atoms at either end of the chain fixed. Atoms adjacent to these terminal atoms are subject to Langevin thermostats set to $T_\text{cold}$. Periodic driving forces of the form $\vec{F}(t) = F_0(\text{cos}\omega t,\text{cos}(\omega t+\phi),0)$ are applied to atoms at the center of the chain, and the steady-state heat flux $J_{L}$ ($J_{R}$) that develops in the minus (plus) $z$-direction is measured.}
\end{figure}

The second model (referred to below as Model 2) we consider is a polyethylene double-helix \cite{ethan2023chain,ethan2023quantifying} with more realistic bond, angle, dihedral, and inter-chain potentials, and thus can be analyzed only using numerical molecular dynamics only. The Hamiltonian is given by\begin{equation}
\begin{split}
H_2 & = \sum_{j=1}^{2N}\frac{\Vec{p}_j^2}{2m} + \sum_{j=1}^{2N-2} k_b (l_j-l_0)^2 + \sum_{j=1}^{2N-4} k_\theta (\theta_j-\theta_{0})^2 + \\
  & \sum_{j=1}^{2N-6} \sum_{n_j}^4 \frac{C_n}{2} \left[ 1+ (-1)^{n_{j-1}} {\rm{cos}} ( n_j \phi_j) \right] + \\ & \sum_{j=1}^{2N} \sum_{i \neq j} 4 \epsilon_{ij} \left[  (\frac{\sigma}{r_{ij}})^{12} - (\frac{\sigma}{r_{ij}})^{6}  \right],
\end{split}
\label{Hmol}
\end{equation} where $N$ is the polymerization of a single chain of the double-helix, $\dot{\vec{x}}_j$ is the velocity of atom $j$, $l_j$,$\theta_j$, and  $\phi_j$ are the values of the $j$th bond-length, angle, and dihedral respectively, and $r_{ij}$ are the inter-atom distances. The parameters $k_{b}$ and $k_{\theta}$ are the spring constants for the bonds and angles, and $C_n$ are constants that define the dihedral potentials. The $\sigma$ and $\epsilon_{ij}$ are Lennard-Jones parameters to account for non-bonded interactions  (such as between the strands in opposite chains). As in Ref. \citen{Hadi} the above parameters were chosen to fit observed physical properties \cite{Martin1998,Wick2000}.

Figure 1 shows the construction used in this study, where the chain constructed either according to model (1) or (2) above is oriented along the $z$-direction. A section in the middle of the chain (a single central atom in the case of (1) and the two central atoms in each strand in the case of (2)) is subjected to a coherent driving force in the plane perpendicular to the chain length
\begin{equation}
\begin{split}
\vec{F}(t) = F_0(\text{cos}\omega t,\text{cos}(\omega t+\phi),0),
\end{split}
\label{driving}
\end{equation}
where $F_0$ is the amplitude of the driving force, $\omega$ is its frequency, and $\phi$ is a phase lag between the $x$ and $y$-components.

We defined the site-resolved kinetic temperature $T_j$ of atom $j$ by equipartition as $T_j=\frac{2}{3k_B}\langle E_{k,j}\rangle=\frac{m}{3k_B}\langle \dot{\vec{x}}_j^2\rangle$. In most of the calculations described below, the temperature at the two chain ends was taken as $T_{cold}=0~\mathrm{K}$. This choice is easily implemented by subjecting the segments at the chain ends to damping forces of the form $-\gamma m \dot{\vec{x}}_j$ where $\gamma$ is a damping constant. Once the system reaches steady-state, the heat currents flowing in either direction are then obtained from
\begin{subequations}\label{current1}\begin{align}
J_{L} &= \sum_{j\in S_L} \gamma m \dot{\vec{x}}_j^2,\\
J_{R} &= \sum_{j\in S_R} \gamma m \dot{\vec{x}}_j^2,
\end{align}\end{subequations} where $S_L$ and $S_R$ are groups of atomic sites at the left and right chain ends respectively. These groups were taken large enough (10-20 sites) to insure that the temperature fell to zero by the ends of the chain (see Fig. 2). Interestingly, in one numerical test reported below, we found that the heat rectifying behavior still persists when $T_\text{cold}=300$ K.

This heat flux observable defined in  Eq. (\ref{current1}) can be obtained numerically via standard velocity Verlet integration of classical trajectories. In the case of model (1) this is achieved via an in-house code using integration timestep $\Delta t=0.25~\mathrm{fs}$, while for (2) it is achieved using the LAMMPS code (the version corrected for heat flux \cite{LAAMPS1,LAAMPS2}) using integration timestep $\Delta t=1.25$ fs. 

In the case of the harmonic model (1), the heat flux can be evaluated analytically as well using the methods discussed in Ref. \cite{Feng2025Nuclear}. In particular, the equations of motion are derived from Eq. (\ref{harmonic}), supplemented by an external driving force $\Vec{F}_j$ on each atom j and the dissipation $\Vec{f}_j=-m\gamma_j\Dot{\Vec{x}}_j$
    \begin{equation}
        \begin{aligned}
              &m\Ddot{\Vec{x}}_1=-\textbf{K}_{1,2}(\Vec{x}_1-\Vec{x}_{2})-\textbf{K}_{1,3}(\Vec{x}_1-\Vec{x}_{3})-\textbf{K}_{1,4}(\Vec{x}_1-\Vec{x}_{4})\\
              &~~~~~~~~~~~+\Vec{F}_1-m\gamma_1 \Dot{\Vec{x}}_1,\\
              &m\Ddot{\Vec{x}}_j=-\sum_{\nu=1,2,3}\left[\textbf{K}_{j-\nu,j}(\Vec{x}_j-\Vec{x}_{j-\nu})+\textbf{K}_{j,j+\nu}(\Vec{x}_j-\Vec{x}_{j+\nu})\right]\\
              &~~~~~~~~~~~+\Vec{F}_j-m\gamma_j \Dot{\Vec{x}}_j,\\
              &\quad \quad \quad \quad \vdots\\
              &m\Ddot{\Vec{x}}_N=-\textbf{K}_{N-1,N}(\Vec{x}_N-\Vec{x}_{N-1})-\textbf{K}_{N-2,N}(\Vec{x}_{N}-\Vec{x}_{N-2})\\
              &~~~~~~~~~~~-\textbf{K}_{N-3,N}(\Vec{x}_{N}-\Vec{x}_{N-3})+\Vec{F}_N-m\gamma_N \Dot{\Vec{x}}_N,\\
          \end{aligned}
    \end{equation}

can be cast in the compact form 
    \begin{equation}\label{eq:compact}
        m\Ddot{\mathbb{X}}=\mathbb{K}\cdot \mathbb{X}-m\mathbb{\Gamma}\cdot \Dot{\mathbb{X}}+\mathbb{F},
    \end{equation}
    where

    \begin{equation}
        \mathbb{X}=\left (\Vec{x}_1,\Vec{x}_2, \Vec{x}_3,...,\Vec{x}_N \right )^T,
    \end{equation}
    
    \begin{equation}
        \mathbb{\Gamma}=\mathrm{Diag}\left [\gamma_1,\gamma_1,\gamma_1,...,\gamma_j,\gamma_j,\gamma_j,...,\gamma_N \right],
    \end{equation}
and
    \begin{equation}
        \mathbb{F}=\left[ \Vec{F}_1, \Vec{F}_2, \Vec{F}_3,...,\Vec{F}_N \right]^T.
    \end{equation} The expression for the strength tensor $\mathbb{K}$ can be found in the appendix section I. Note that in our construction, the elements of $\mathbb{\Gamma}$ are nonzero only for the indices $j \in S_L \cup S_R$, and the elements of $\mathbb{F}$ are only nonzero for the central driven atom(s), where the respective values have been specified above.

The steady state reached while driving with frequency $\omega$ is described by the Fourier transform of Eq.~(\ref{eq:compact})
    \begin{equation}        
    -m\omega^2\Tilde{\mathbb{X}}=\mathbb{K}\cdot\Tilde{\mathbb{X}}-im\omega\mathbb{\Gamma}\cdot \Tilde{\mathbb{X}}+\Tilde{\mathbb{F}},
    \end{equation} where the tildes denote Fourier conjugates in frequency space, e.g $\tilde{\mathbb{X}}=\tilde{\mathbb{X}}(\omega).$ The solution to Eq.~\ref{eq:compact} is
\begin{equation}
\label{equation:FT_X}
    \Tilde{\mathbb{X}}(\omega)=\tilde{\mathbb{G}}(\omega)\Tilde{\mathbb{F}}(\omega)
\end{equation}
where $\tilde{\mathbb{G}}$ is the Green’s function
    \begin{equation}\label{greens}
        \tilde{\mathbb{G}}(\omega)=[-\mathbb{K}+im\omega\mathbb{\Gamma}-m\omega^2\mathbb{1}]^{-1}.
    \end{equation}
Thus, using Eq. (\ref{driving}) and  Eq. (\ref{greens}) in Eq. (\ref{equation:FT_X}) and then inverting the Fourier transform, the motion of each atom $j$ is obtained
    \begin{equation}
    \label{equation:XY}
        \begin{aligned}
            x_j(t)&=(A_je^{i \omega t}+A_j^*e^{-i \omega t})\Tilde{F}_0\\
            y_j(t)&=(B_je^{i \omega t}+B_j^*e^{-i \omega t})\Tilde{F}_0\\
            z_j(t)&=(C_je^{i \omega t}+C_j^*e^{-i \omega t})\Tilde{F}_0.\\
        \end{aligned}
    \end{equation} Using Eq. (\ref{equation:XY}) in Eq. (\ref{current1}) and averaging over time, we obtain
\begin{subequations}\label{current2}\begin{align}
J_{L} &= \sum_{j\in S_L} \omega^2 m\gamma_j(|A_j|^2+|B_j|^2+|C_j|^2)\tilde{F}_0^2\\
J_{R} &= \sum_{j\in S_R} \omega^2 m\gamma_j(|A_j|^2+|B_j|^2+|C_j|^2)\tilde{F}_0^2,
\end{align}\end{subequations} which is the desired analytical result.

The dimensionless ratio $J_{R}/J_{L}$, computed using either of the methods outlined above, is a measure of thermal rectification; in what follows we report this quantity for both models as a function of the parameters that characterize the driving force in Eq. (\ref{driving}). 

\section{Results}

 As outlined in the introduction, heat transport in chiral environment generates excitations of atomic motions with angular momentum (“chiral phonons”) whose sign depends on the direction of heat current and the molecular handedness. This behavior, together with the thermal distribution (defined as $\mathrm{(2/3k_B)\langle E_K} \rangle$ where $\mathrm{E_K}$ is the atomic kinetic energy) along the chain of Fig. 1 is demonstrated in Fig. 2 for (a-b) Model 1 and (c-d) Model 2. On the one hand, focusing on the temperature profile (Fig. 2(a) and 2(c)), we clearly observe that when the driving force is circular (in this case clockwise), the temperature on the left-hand side is significantly higher than the right-hand side. The same effect holds for the heat flux. The higher temperature to the left of the driving point at Z=0 indicates a larger energy flux going leftward (the heat flux ratio was calculated to be ($J_L/J_R$=0.021 for Fig.~2a and 0.34 for Fig.~2c).  On the other hand, linear driving along a certain direction (in this case along $x=y$) does not have a directional preference, as the orange lines are roughly symmetric about the center of the chain. For the angular momentum observable (Fig.~2(b) and 2(d)), we observe the opposite angular momentum polarization on the opposite side of the chain, a result that persists regardless of the driving phase. This observation is in agreement with our recent study of the angular momentum observable in response to white-noise driving \cite{Feng2025Nuclear}, and likewise can be explained as a linear-angular momentum locking effect.

\begin{figure}
\includegraphics[width=250 pt]{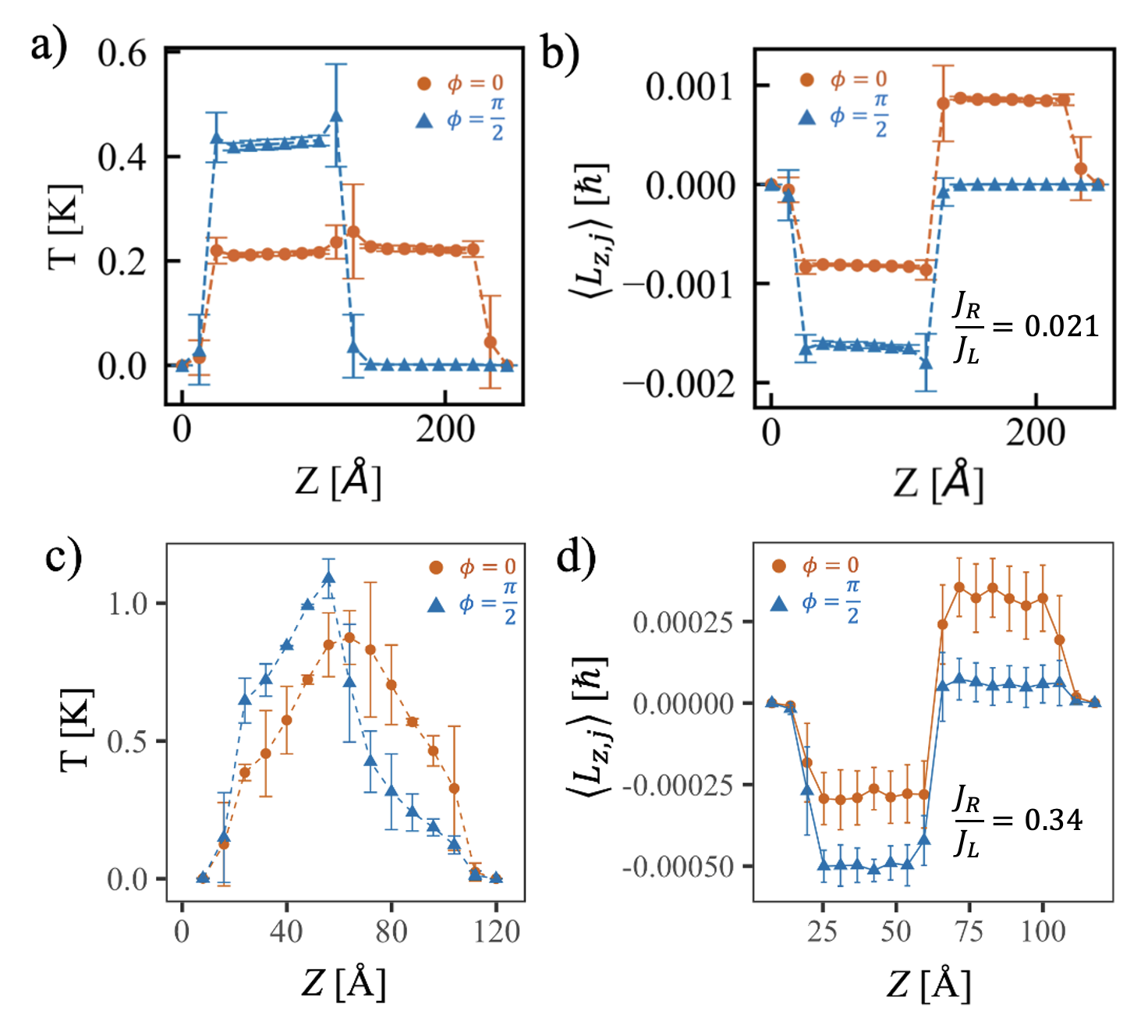}
\label{fig:2} 
\caption{Spatial profiles of temperature and nuclear angular momentum under phase-shifted driving:
  (a–b) Harmonic chain (model~1, $N=201$) driven at $\omega=530.9~\mathrm{cm^{-1}}$ ($15.9$~THz).
  (c–d) Polyethylene double helix (model~2, $N=98$; 15 left-handed twists) driven at
  $\omega=955~\mathrm{cm^{-1}}$ (28.6 THz).
  All simulations use a driving force with amplitude $F_{0}=0.5~\mathrm{kcal}/ \mathrm{mol}^{-1}$\AA$^{-1}$ and phase $\phi=0$ (orange circles) or
  $\phi=\pi/2$ (blue triangles). $Z$ denotes position along the chain axis. (a,c) The time-averaged local kinetic temperature $T(Z)$. (b,d) Site-resolved nuclear
  angular momentum $\langle L_{z,i}\rangle/\hbar$. Error bars indicate variability within a block of atoms.}
\end{figure}

The directional energy transport characterizing these systems is more clearly seen in Figures 3 – 6. Fig. 3 and 4 show the ratio  $J_\text{R}/J_{\text{L}}$ as a function of the driving frequency for models 1 and 2, respectively. The strong frequency dependence of the rectification phenomenon follows directly from the solution of Eq.~22, which yields the density of phonon states as well as the phonon polarization in the phonon band structure. A detailed plot of the density of states as well as the phonon band structure for the single chain is shown in Ref. \cite{Feng2025Nuclear}. The currents are larger in one direction than in the other when the polarization of the phonon is strongly coupled with the group velocity (so that excitation with a certain polarization and frequency of the system will lead to the phonon generation with certain direction of the group velocity). These results were calculated for the driving phase  $\phi=\pi/2$ and amplitude $F_{0}=0.5~\mathrm{kcal}~ \mathrm{mol}^{-1}$\AA$^{-1}$). Note that Fig. 3 presents numerical results obtained for the full Model 1 as well as results obtained from the analytical solution of the harmonic approximation. The excellent agreement seen indicates the validity of the harmonic approximation for the chosen small amplitude driving.

\begin{figure}
\includegraphics[width=140 pt]{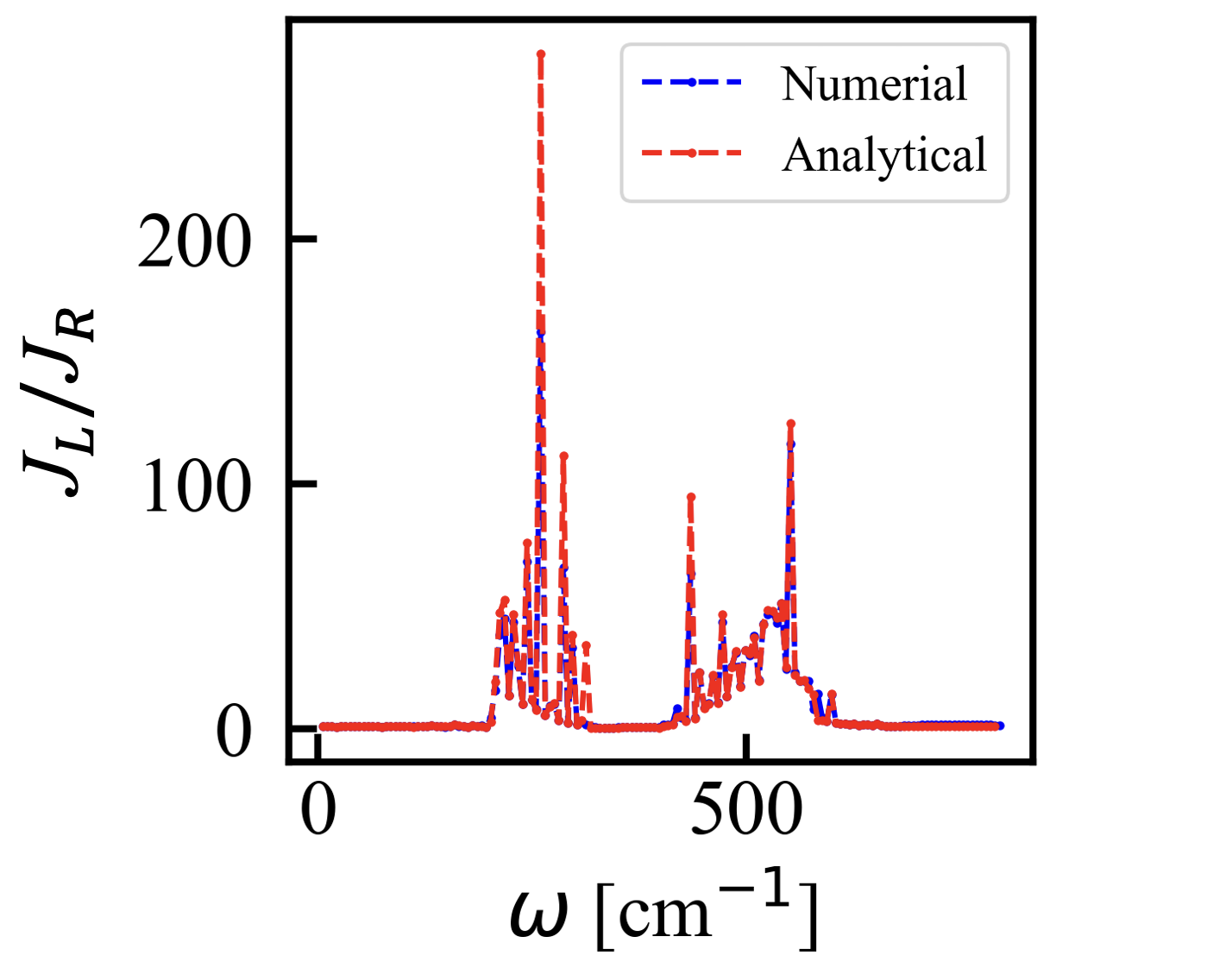}
\label{fig:3} 
\caption{Analytical and numerical results for the rectification ratio $J_{L}/J_{R}$ plotted against driving frequency $\omega$ for the harmonic chain (Model 1, $N=201$). Here the $y$-axis ratio is $J_L/J_R$ rather than $J_R/J_L$ in order to show the magnitude of the peaks located at $200-600~\mathrm{cm}^{-1}$.}
\end{figure}

Remarkably, we find that at various driving frequencies, both ratios $J_\text{R}/J_{\text{L}} \gg 1$ and $J_\text{R}/J_{\text{L}} \ll 1$ are observed, and such values can differ from the unity by a factor of $10^2$--$10^3$. We note the spectral quality of this rectifying behavior: the same structure will favor heat transport in different directions depending on the frequency. Furthermore, we find that changing either the helical enantiomer (results not shown) or the polarization of the driving (see discussion in Fig.~4 below) will reverse the directional preference at a particular frequency (i.e. $J_\text{R}/J_{\text{L}} \rightarrow J_\text{L}/J_\text{R}$).

Interestingly, for the case of the double helical Model 2 with chain length $N=98$, a significant rectification is observed (Fig. 4(a-b)) only in the highly twisted (15 twists) structure, while for a mildly twisted (3-twists) structure  $J_\text{R}/J_{\text{L}}$ is close to 1 at all frequencies.

\begin{figure}
\includegraphics[width=240 pt]{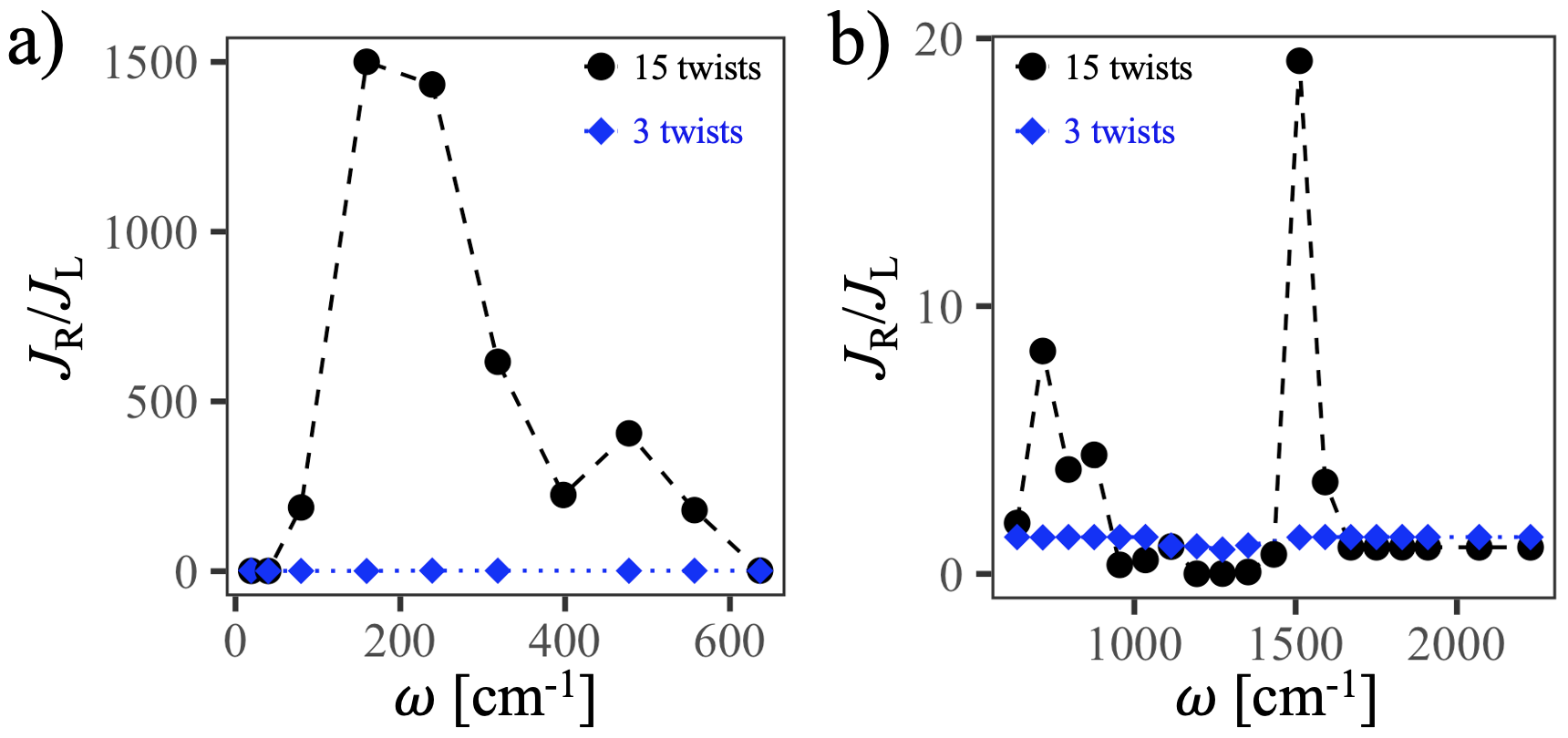}
\label{fig:4} 
\caption{Numerical results for $J_{R}/J_{L}$ plotted against driving frequency $\omega$, for the polyethylene double-helix (Model 2, $N=98$) containing 15 twists (black circles) and 3 twists (blue diamonds). Due to large variations in the range of $J_{R}/J_{L}$ for Model 2 across the frequency spectrum, results for the low-frequency range (a) and high-frequency range (b) are displayed separately.}
\end{figure}

\begin{figure}
\includegraphics[width=240 pt]{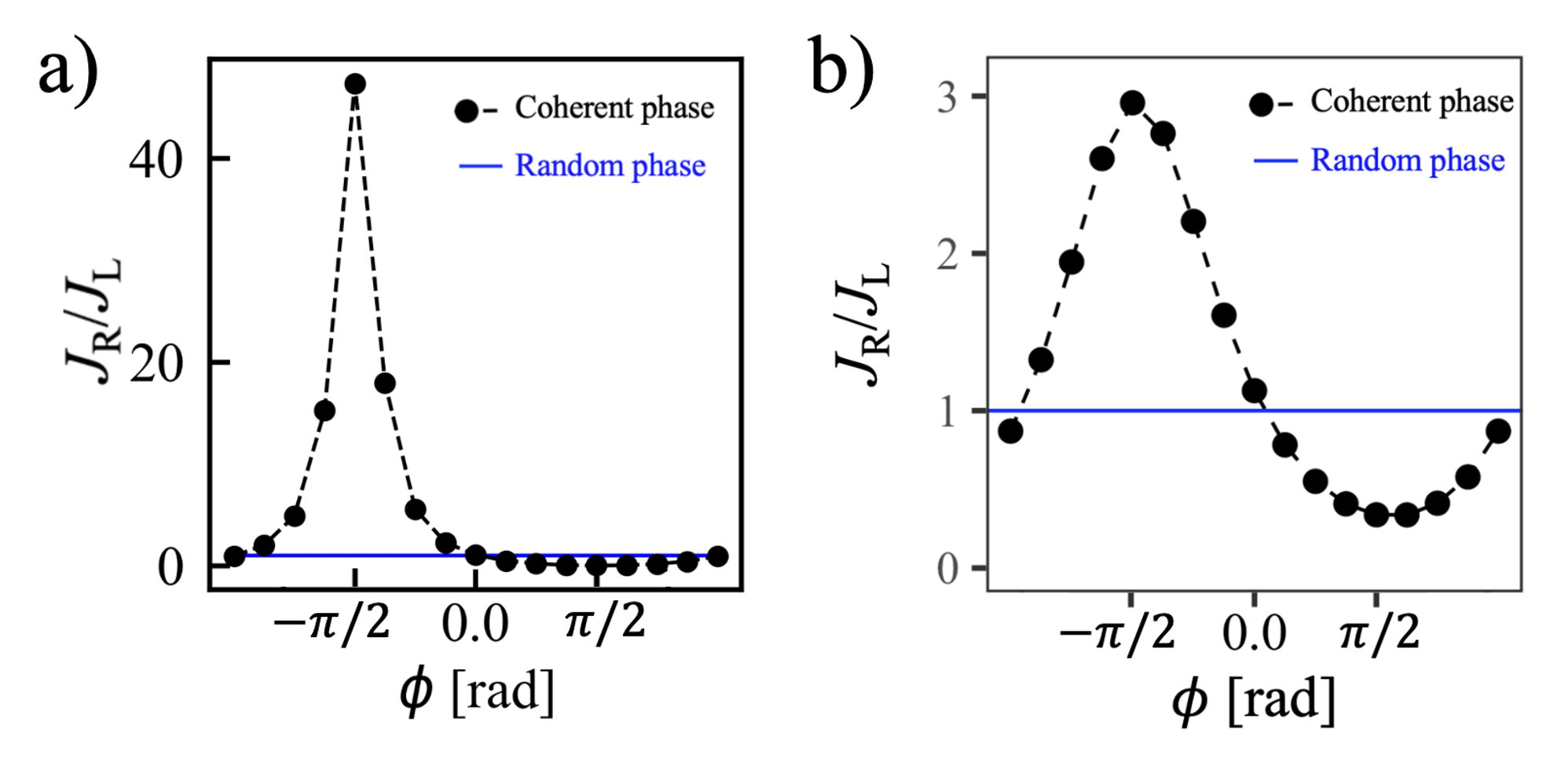}
\label{fig:5} 
\caption{The rectification ratio $J_{R}/J_{L}$ as a function of phase $\phi$ for (a) the harmonic chain (Model 1, $N=201$) driven at $\omega=530.9~\mathrm{cm^{-1}}$ ($15.9$~THz), and (b) the polyethylene double-helix (Model 2, $N=98$) driven at $\omega=955~\mathrm{cm^{-1}}$ ($28.6$~THz). The blue line shows the non-result ($J_{R}/J_{L}=1$) for driving without phase coherence.}
\end{figure}

Fig.~5  show similar results for $J_R/J_L$, now plotted against the driving phase $\phi$ (see Eq.~14) at fixed frequency.  Results are shown for the two models in panels (a) and (b) respectively. We observe the greatest difference from unity when the driving phase is equal to $\pm \pi/2$, and that reversing the phase ($\phi \rightarrow -\phi$) results in the reciprocal rectification ratio ($J_\text{R}/J_{\text{L}} \rightarrow J_\text{L}/J_\text{R}$). The peaks at $\pm \pi/2$ driving phase reflect the nature of circular motion of chiral phonons. 

Using the analytical results obtained for the harmonic approximation to Model 1 we can also examine the combined effect of the driving frequency ($\omega$) and phase ($\phi$). Fig.~6 shows a heat map showing the rectification displayed against both $\omega$ and $\phi$. As for  Fig.~3, the phase dependence of the rectification effect is related to the structure of chiral phonon modes and the frequency dependence results from both the polarization of the phonon band structure and the phonon density of states. 

\begin{figure}
\includegraphics[width=140 pt]{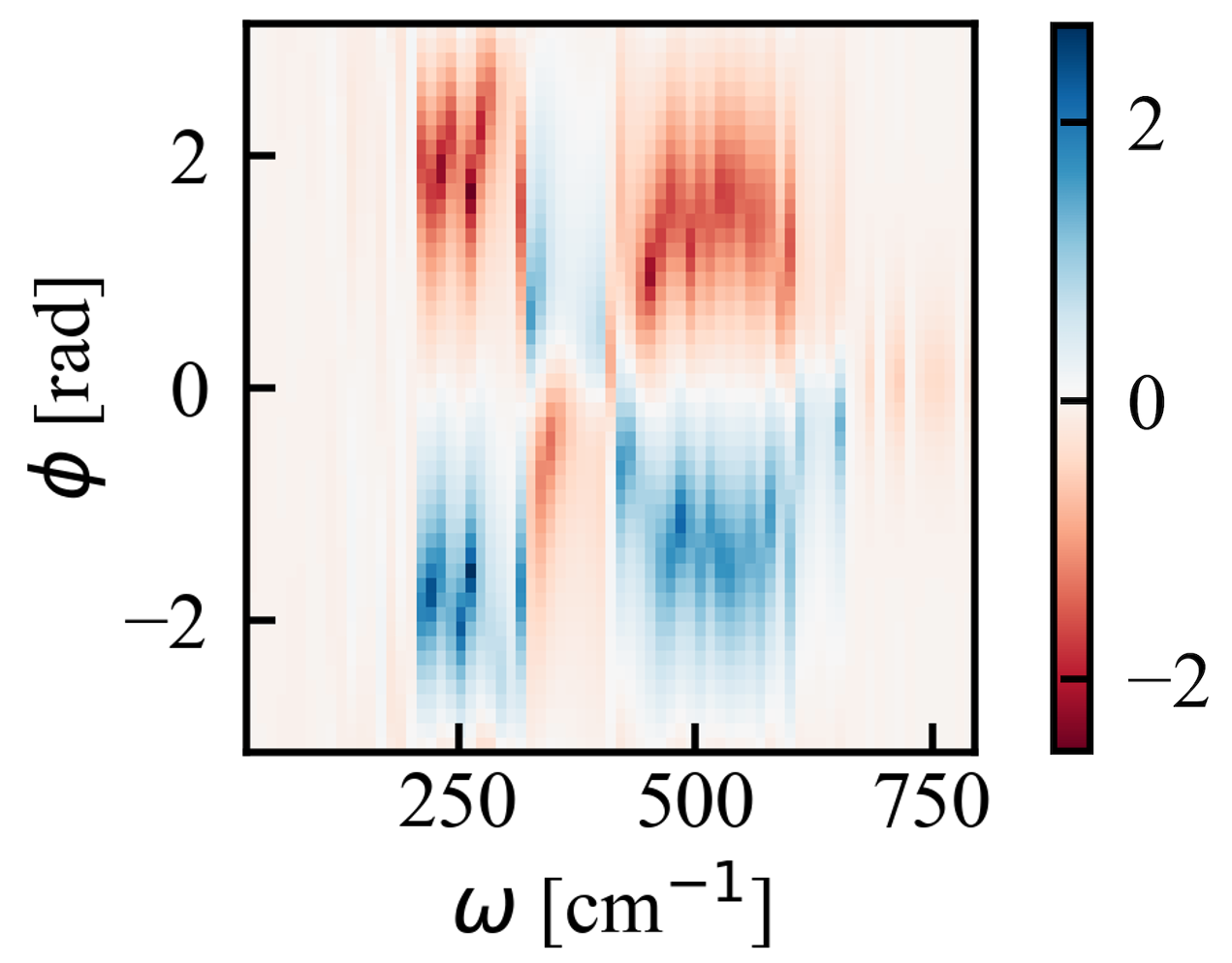}
\label{fig:6} 
\caption{Heatmap of $\log_{10}(J_R/J_L)$ with respect to the driving frequency $\omega$ and the driving phase $\phi$.}
\end{figure}

Finally, to address the extent to which the findings outlined above can be expected to influence room-temperature experiments, we conducted sample Langevin molecular dynamics simulations with $T_\text{cold}=300$ K according to the methods described in Ref. \cite{Feng2025Nuclear}. We used the $N=98$ polyethylene double-chain model containing 15 twists described above, and the driving parameters were set to $\phi=\pi/2$, $\omega=79.5$ cm$^{-1}$ (2.38 THz), and $F_{0}=4.0~\mathrm{kcal}/ \mathrm{mol}^{-1}$\AA$^{-1}.$ We performed five independent simulations of length 10 ns, and we obtain a ratio of $J_\text{R}/J_{\text{L}}=32.5\pm3.1.$ This result strongly suggests that the phenomenon observed in this study is relevant to room-temperature experiment.

\section{Conclusion}
In this work, we analytically and computationally study the heat rectification phenomenon in a chiral system with phase-controlled driving. We examined two structures: a single helical chain (studied without and with the harmonic approximation) and a polyethylene double helix. We found significant asymmetric heat flux under circularly polarized driving at the center of the system. The strong frequency dependence of this phenomenon allows us to modify the rectification effect by tuning the driving frequency, sometimes even switching the direction of the heat transfer, which is also achievable by inverting the polarization of the circular driving. The effect is highly phase dependent, as it decreases when the driving becomes more linearly polarized and is maximal when the driving is circularly polarized in the plane perpendicular to the chain direction. The effect is also correlated with the structural chirality, such that the effect is much weaker (orders of magnitude) in less chiral systems (a polyethylene double helix with 3 twists) than in more chiral systems (a polyethylene double helix with 15 twist). The agreement between analytical results and molecular dynamics in the harmonic model ensures that the effect does not hinge on numerical artifacts. The effect is robust enough that in a double helix system with realistic interaction potentials and room-temperature baths, we still get a remarkable heat flux ratio (e.g. $J_R/J_L\sim30$ at $\omega=79.5~\mathrm{cm}^{-1}$), which is a large effect and indicates the feasibility of experimental realization, perhaps by local driving using circularly polarized light.

Future work will focus on quantum-regime extensions with electron-phonon and electron-spin coupling. Although in our simulations the drive is applied at the chain center as a computational convenience, local circularly polarized driving force can be applied to one end of the molecular chain by introducing near-field circularly polarized light (CPL) and with the form 
\begin{equation}
    \Vec{F}=q\Re (\Vec{E}e^{i\omega t})
\end{equation}
which is effectively equivalent to the  external force in our computational setup above.  Changing the enantiomer will also lead to a similar reversal as changing the polarization of the driving. This chiral rectifier/diode effect gives rise to the possibility of building controllable molecular energy transport networks, and possibly CPL catalysis in chiral systems.

\section*{ACKNOWLEDGMENTS}
This paper is submitted in honor of Prof. YiJing Yan and his seminal contribution to the research of molecular dynamical processes. The research of A.N. is supported by the National Science Foundation award number 2451953. The research of J.E.S. is supported by the Air Force Office of Scientific Research under award number FA9550-23-1-0368. E.A acknowledges the support of the Massachusetts Institute of Technology Department of Chemistry and Department of Chemical Engineering. We thank Claudia Climent for many helpful discussions.

\section*{AUTHOR DECLARATION}
\subsection*{Conflict of Interest}
The authors have no conflicts to disclose.

\section*{DATA AVAILABILITY}

The data that support the findings of this study are available from corresponding authors upon reasonable request. Structures corresponding the double-chain model examined above are available at \url{https://github.com/eabes23/polymer_twist/}.

\section*{APPENDIX: MODEL PARAMETERS}
In Model 1, the strength constant for the coupling term is $K=500~\mathrm{N/m}$, the bond length for the nearest coupling is $a_1=1.5~$\AA, $a_2=2.7~$\AA, $a_3=4.0~$\AA. The mass of the atom is $m=12~\mathrm{AMU}$.
\begin{widetext}
The strength tensor can be expressed as
\begin{equation}
    {\scriptsize
    \begin{aligned}
        &\mathbb{K}=\\
        &\begin{pmatrix}
            -\textbf{K}_{1,2}-\textbf{K}_{1,3}-\textbf{K}_{1,4} &\textbf{K}_{1,2}  & \textbf{K}_{1,3} & \textbf{K}_{1,4} & 0 &...& 0 \\
            \textbf{K}_{1,2} & -\textbf{K}_{1,2}-\textbf{K}_{2,3}-\textbf{K}_{2,4}-\textbf{K}_{2,5} & \textbf{K}_{2,3} & \textbf{K}_{2,4} & \textbf{K}_{2,5} &...& 0 \\
            \textbf{K}_{1,3} & \textbf{K}_{2,3} & -\textbf{K}_{1,3}-\textbf{K}_{2,3}-\textbf{K}_{3,4}-\textbf{K}_{3,5}-\textbf{K}_{3,6} & \textbf{K}_{3,4} & \textbf{K}_{3,5} &...& 0 \\
            \textbf{K}_{1,4} & \textbf{K}_{2,4} & \textbf{K}_{3,4}& -\textbf{K}_{1,4}-\textbf{K}_{2,4}-\textbf{K}_{3,4}-\textbf{K}_{4,5}-
            \textbf{K}_{4,6}-
            \textbf{K}_{4,7}& \textbf{K}_{4,5}&...& 0 \\
            0& 0& 0 &... &  &  &  \\
            \end{pmatrix},
    \end{aligned}}
    \end{equation}
\end{widetext}

In Model 2, the force field parameters used are $k_b=450$ kcal mol$^{-1}$\AA$^{-2}$, $l_0=1.54$ \AA, $k_\theta=62.1$ kcal mol$^{-1}$rad$^{-2}$, $\theta_0=114^\circ$, $C_1=1.4110$ kcal mol$^{-1}$, $C_2=-0.2708$ kcal mol$^{-1}$, $C_3= 3.143$ kcal mol$^{-1}$, $C_4=0.0$ kcal mol$^{-1}$, $\sigma=3.95$ \AA, and $\epsilon_{ij}=0.0912$ kcal mol$^{-1}$ (for $|\vec{r}_i-\vec{r}_j|<14 \text{\AA}, =0$ otherwise). The particle mass is taken $m=14$ AMU, modeling each CH$_2$ bead of the polyethylene chain as a single interaction site.

\setcounter{figure}{0}
\renewcommand{\figurename}{FIG.}
\renewcommand{\thefigure}{S\arabic{figure}}

\setcounter{equation}{0}
\renewcommand{\theequation}{S\arabic{equation}}

\bibliography{reference}
\end{document}